\documentclass[aps,prb,twocolumn,groupedaddress]{revtex4-1}
\usepackage{graphicx}
\usepackage{dcolumn}
\usepackage{bm}
\usepackage{amssymb}
\usepackage{amsmath}
\usepackage{color} 

\begin{document}


\title{Polymer effects on K\'arm\'an Vortex: Molecular Dynamics Study}



\author{Yuta Asano}
\email[]{yuta.asano@issp.u-tokyo.ac.jp}
\author{Hiroshi Watanabe}
\author{Hiroshi Noguchi}
\affiliation{The Institute for Solid State Physics, The University of Tokyo, Kashiwanoha 5-1-5, Kashiwa, Chiba 277-8581, Japan}

\date{\today}

\begin{abstract}
We investigated the K\'arm\'an vortex behind a circular cylinder in a polymer solution by a molecular dynamics simulation. The vortex characteristics are distinctly different for short and long polymers. The solution with the long polymer exhibits a reduction in the vortex shedding frequency and broadening of the lift coefficient spectrum. On the other hand, the characteristics of the short-polymer solution are almost same as those of the Newtonian fluid. These facts are consistent with the experiments. Because the distributions of the gyration radius and the orientational order of the long-polymer solution are highly inhomogeneous in the flow field, we conclude that the extensional property of the polymer plays an important role in changing the flow characteristics.
\end{abstract}


\maketitle 

%
%
%
%
%
\section{INTRODUCTION}
The addition of a small amount of polymer into a Newtonian fluid changes its behavior drastically, such as turbulent drag reduction~\cite{gadd65, gadd66a, gadd71} which is so-called Toms effect, and the change in vortex structure in K\'arm\'an vortex street.~\cite{gadd66b} Since the turbulence and vortices can be found ubiquitously in industrial flows, adding polymers has been attracting attention from energy saving, environmental protection, and so forth. Particularly, a tremendous amount of effort has been devoted to investigate the Toms effect because the Toms effect can save energy consumption for transportation of fluid.~\cite{sw00, wm08, hdc17} Despite its importance, the detailed mechanism of drag reduction by polymers remains unclear. There are mainly two difficulties; one is the phenomenon of the turbulence itself, and the other is incomplete understanding of the polymer rheology in flow. Various scales of eddies exist in turbulent flow. Although the size of the polymer is much smaller than the smallest eddy size in turbulent flow, interactions between the polymer and the vortical motion near a wall would play an important role in the drag reduction.~\cite{myc03} Therefore, in order to elucidate the mechanism of the Toms effect, it is necessary to analyze polymer behavior in the vortex. 

Although the flow around a circular cylinder is a fundamental problem of fluid dynamics, understanding the flow characteristics is important because it involves many physical phenomena such as the behavior of drag,~\cite{henderson95} vortex motion,~\cite{pcl82,gerrard66} the characteristics of the Aeolian tone,~\cite{phillips56} and so forth. For the Newtonian fluid,  many experimental and numerical studies have been reported for flows past a circular cylinder.~\cite{bw72,oertel90, williamson96} The flow is characterized by the Reynolds number $Re=\rho D V / \eta$, where $\rho$, $D$, $V$ and $\eta$ are the fluid density, cylinder diameter, inlet velocity and fluid viscosity, respectively. For $Re$ up to $49$, a steady recirculation region consisting of a symmetric pair of vortices is formed behind the cylinder. The length of the recirculation region increases as $Re$ increases. Then, the vortex shedding appears at $Re=49$. In the range $49\le Re \le 140 \textendash 194$, the velocity oscillation near wake region becomes enhanced and the formation length of the vortex decreases as $Re$ increases. In the range $190 < Re < 260$, the transition from the two-dimensional wake to the three-dimensional wake occurs.~\cite{lw98} There are two transitions; one occurs near $Re=180 \textendash 194$, and the other occurs near $Re=230 \textendash 250$. The former is characterized by the appearance of streamwise vortex loops form the three-dimensional structure whose spanwise wavelength is around $3 \textendash 4D$. The latter is characterized by the formation of finer scale structures of the streamwise vortex. In the range $1000 < Re < 200~000$, the shear layer transition occurs. 

As regards the flow of the dilute polymer solution, a number of experimental studies have been conducted to investigate how polymers affect the flow field. Gadd observed the vortex shedding frequency at $Re=240$ for solutions of polyethylene oxide (PEO), guar gum, and polyacrylamide.~\cite{gadd66b} These solutions cause the Toms effect. He found that the frequency of the dilute polymer solution of PEO is lower than that of water, and the frequency decreases with increasing polymer concentration. In the case of guar gum and polyacrylamoide, such changes were not observed. However, in a subsequent study by Kim and Telionis,~\cite{kt89} the reduction in the shedding frequency for the polyacrylamide was confirmed.

Kalashnikov and Kudin observed the vortex shedding frequency in the range $Re<400$ for a 10 ppm solution of PEO.~\cite{kk70} They found that the amount of the reduction in the vortex shedding frequency for the PEO solution depends on the cylinder diameter. They also showed that the critical Reynolds number at which the vortex shedding occurs is reduced by the addition of PEO. An important finding obtained from their experiment is that the frequency of the inelastic fluids, such as guar gum and degraded PEO solution, is greater than that of water. This result suggested that the polymer elasticity was responsible for the change in the vortex shedding frequency.

The relationship between the vortex shedding frequency and the polymer elasticity was first investigated by Usui~{\it et~al.}~\cite{uss80} They showed that the reduction in the vortex shedding frequency of the PEO solution is correlated with the Weissenberg number and the elasticity number. In the experiments of Cadot~{\it et~al.}~\cite{cl99, ck00} and Cressman~{\it et~al.},~\cite{cbg01} it was observed that a solution of low-molecular-weight PEO has no effect on the flow field. Since the molecular weight makes a large contribution to the elongational viscosity, Cressman~{\it et~al.} suggested that the elongational viscosity plays a key role in the change in the flow field.

A few numerical simulations have been conducted for the flow of a dilute polymer solution past a circular cylinder.~\cite{oliveira01, so04, ris10, xbk11, nvm13, xby17} In these simulations, the polymer solution is treated by the constitutive equations, such as the finitely extensible nonlinear elastic (FENE) model. Oliveira showed that the modified FENE Chilcott--Rallison (MCR) model can reproduce almost all the experimental results, such as the reduction in the vortex shedding frequency and the lengthening of the vortex structure.~\cite{oliveira01} The only difference between his simulation results and the experimental results is the critical Reynolds number. Although the experimental results showed that the critical Reynolds number decreases due to addition of the polymer,~\cite{kk70, uss80} Oliveira's simulation showed that the critical Reynolds number increases due to the elasticity. This stabilization effect was confirmed by Sahin and Owens,~\cite{so04} who performed a stability analysis by using both direct numerical simulations and the solution of a generalized eigenvalue problem for MCR fluid. In this regard, experiments on the stabilization of the polymer solution were conducted by Pipe and Monkewtiz for the PEO solution.~\cite{pm06} Their detailed analysis showed that the flow is stabilized by adding the polymer, and they suggested that the stabilization is associated with the high shear rate behind the cylinder. Furthermore, they suggested that the stabilizing effect is counteracted by the shear thinning. A similar suggestion was made by Coelho and Pinho,~\cite{cp03_1,cp03_2} who demonstrated that the shear thinning has the opposite effect to fluid elasticity on fluid stability.

Although previous studies indicate that the elongational properties of polymers play an important role in the vortex shedding, little is known about polymer behaviors in the flow field. This is because it is difficult to trace the motion of individual polymer chains. In order to discuss the effect of the polymer on the flow field, it is necessary to treat the motion of the polymer and solvent as directly as possible.
Mesoscopic hydrodynamic computational methods such as the lattice Boltzmann method (LBM)~\cite{succ01} and the multi-particle collision (MPC) dynamics~\cite{mk99} are powerful tools for analyzing complex flows. In fact, LBM and MPC were applied for analyzing the K\'arm\'an Vortex in the Newtonian fluid.~\cite{hd97_1,hd97_2,lg02,lgi01}
However, taking into consideration of today's computational power, the flow can be treated by directly solving the Newtonian equations of motion.
Hence, we analyzed a flow of dilute polymer solution past a circular cylinder by a molecular dynamics (MD) simulation. An MD simulation of the K\'arm\'an vortex street for a simple liquid has already been conducted by Rapaport~{\it et~al.}~\cite{rc86, rapaport87} Here, two-dimensional MD simulations were conducted to elucidate the mechanism of the change in the flow field by adding polymer chains.
\section{METHOD}
\subsection{Model}
A solvent particle is modeled by a monoatomic molecule whose interparticle interaction is given by the Weeks--Chandler--Andersen (WCA) potential~\cite{wca71}:
\begin{eqnarray}
  u_{\rm WCA}(r) &=&
  \left\{
  \begin{array}{ll}
    4\epsilon \left[\left(\dfrac{\sigma}{r}\right)^{12} - \left(\dfrac{\sigma}{r}\right)^{6} + \dfrac{1}{4}\right] & \left(r\le 2^{\frac{1}{6}}\sigma\right),\\%
    0 & \left(r > 2^{\frac{1}{6}}\sigma\right),
  \end{array}
  \right.\label{eq:wca}
\end{eqnarray}
where $r$ denotes the interparticle distance, and $\epsilon$ and $\sigma$ represent the energy and the length scales, respectively. We refer to the monoatomic molecule whose interparticle potential is given by Eq.~(\ref{eq:wca}) as a WCA particle.

A linear polymer molecule is described by the Kremer--Grest model.~\cite{kg90} The polymer molecule consists of $N_{\rm s}$ WCA particles connected by the FENE potential. The interparticle interaction between segments except for the neighboring segments along a polymer chain is given by Eq.~(\ref{eq:wca}). The interaction between the neighboring segments is given by the following equations:
\begin{eqnarray}
  u_{\rm intra}(r) &=& u_{\rm WCA}(r) + u_{\rm FENE}(r),\\
  u_{\rm FENE}(r) &=&
  \left\{
  \begin{array}{ll}
    -\frac{1}{2}KR_{0}^2 \ln\left(1 - \left(\dfrac{r}{R_{0}}\right)^2\right) & (r\le R_{0}), \\
    \infty & (r > R_{0}),
  \end{array}
  \right.
\end{eqnarray}
where $K$ and $R_{0}$ denote the strength of the interaction and the equilibrium length, respectively. We have used $K=30\epsilon/\sigma^2$ and $R_{0}=1.5\sigma$ throughout this paper.  All particles have an identical mass $m$. Hereafter, physical quantities are measured in units of energy $\epsilon$, length $\sigma$, and time $\tau_0= \sigma\sqrt{m/\epsilon}$. We refer to the fluid without polymers as the reference liquid.

\begin{figure}
\includegraphics{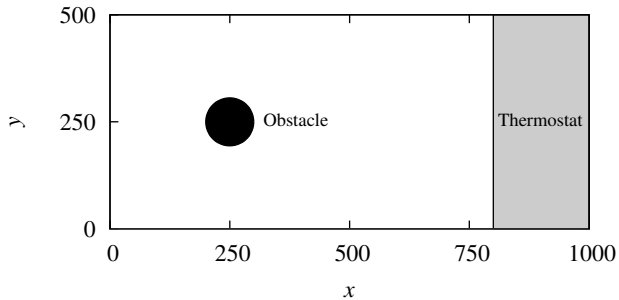}
\caption{\label{fig:simulation_box}Schematic view of the computational domain. The cylinder indicated in a black circle is located at $(x,y)=(250, 250)$ with diameter $D=100$. The Langevin thermostat is employed in the region indicated in gray $(800\le x \le 1000)$.}
\end{figure}

A simulation box is shown in Fig.~\ref{fig:simulation_box}. The system is a rectangle with dimensions $L_x \times L_y$, where $L_{x}=1000$ and $L_{y}=500$. The periodic boundary condition is taken for both directions. The filled circle in Fig.~\ref{fig:simulation_box} denotes a cylindrical obstacle. To satisfy the no-slip boundary condition, the cylinder consists of the WCA particles whose positions are fixed.~\cite{kbw89} The cylinder with a diameter of $D=100$ is located at $(x,y)=(250, 250)$.

In order to impose the uniform inlet velocity $V$, a Langevin thermostat is employed in the left region (the gray region in Fig.~\ref{fig:simulation_box}). The Langevin equation is given by
\begin{eqnarray}
  \dfrac{{\rm d} {\bm v}_i }{{\rm d} t} &=& {\bm F}_i -\zeta \left({\bm v}_i - {\bm V}\right) + {\bm R}_i,
\end{eqnarray}
where $\zeta$ is the friction coefficient, and ${\bm R}_i$ denotes the white noise satisfying $\left\langle R_{i\alpha}(t)R_{j\beta}(t') \right\rangle=2\zeta k_{\rm B}T\delta_{ij}\delta_{\alpha\beta}\delta(t-t')~(\alpha, \beta\in \{x, y\})$. $R_{i, \alpha}$ represents the $\alpha$ component of the white noise. ${\bm F}_i$ denotes the total internal force acting on the $i$th particle. Since we control the velocity ${\bm v}_i$ of $i$th particle via $({\bm v}_{i}-{\bm V})$, the Langevin thermostat produces the uniform velocity ${\bm V}=V {\bm e}_{x}$, where ${\bm e}_{x}$ is the unit vector in the $x$ direction. The temperature is set to $k_{\rm B}T=1$, where $k_{\rm B}$ is the Boltzmann constant. Because the sound speed of the reference liquid $v_{\rm {sound}} \simeq 6$, the velocity of the inlet is less than $3$ to prevent a shock wave.

At the initial states, $N_{\rm p}$ polymer chains are randomly distributed in the simulation box.
The solvent particles and the segments are given initial velocities according to the Maxwell velocity distribution whose average velocity is given by $V {\bm e}_{x}$. The total number of particles $N_{\rm total}=N_{\rm WCA}+N_{\rm s}N_{\rm p}$ is $408~482$. The density $\rho=0.83$ of the system is kept constant. Short- and long-polymer chains are considered: $N_{\rm s}=10$ and $N_{\rm s}=100$. The mole fractions of the polymer are $\phi=0.024$, $0.043$, $0.085$, and $0.107$ in each type of polymer. According to the Flory theory,~\cite{bgm13} the overlap concentration of the polymer having $N_{\rm s}=100$ is approximately $0.1$, which corresponds to a mole fraction of $0.12$. Therefore, in our simulation, the concentration of the polymer is less than the overlap concentration.

The characteristics of the vortex shedding are evaluated by the Strouhal number given by the following equation:
\begin{eqnarray}
  St &=& \frac{f_{\rm ch}D}{V}, \label{eq:st}
\end{eqnarray}
where $f_{\rm ch}$ is the characteristic frequency of the vortex shedding. When the vortex shedding appears, a periodic lift force $F_{\rm L}$ is applied to the cylinder in the $y$ direction. The dimensionless force acting on the cylinder is defined by $C_{\rm L}=2F_{\rm L}/(\rho D V^2)$. We adopt the frequency of the $C_{\rm L}$ as the characteristic frequency $f_{\rm ch}$ of the vortex shedding.

We perform MD simulations using the velocity--Verlet algorithm up to $12~000~000$ steps with a time step of $0.004$.

\subsection{Viscosity}
\begin{figure}
\includegraphics{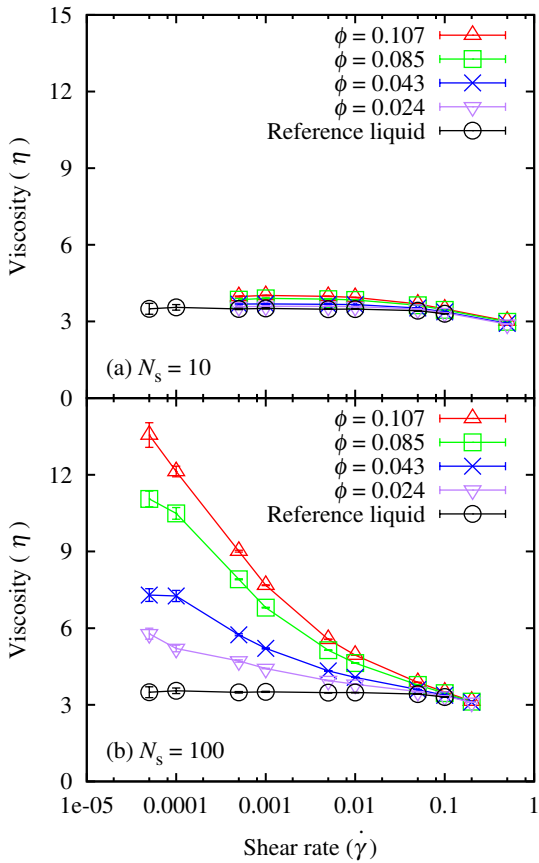}
\caption{\label{fig:viscosity}The shear rate $\dot{\gamma}$ dependence of the viscosity $\eta$
at mole fractions $\phi=0$, $0.024$, $0.043$, $0.085$, and $0.107$ for (a) $N_{\rm s}=10$ and (b) $N_{\rm s}=100$. 
The circles denote the viscosity of the reference liquid ($\phi=0$).}
\end{figure}

To estimate the Reynolds number, we determined the value of the viscosity. We adopt the Lees--Edwards boundary conditions to achieve shear flow.~\cite{le72} The simulation box is a square with side length $500$. The total number of particles is $N_{\rm total}=207~500$.  The shear rate $\dot{\gamma}$ is applied in the $x$ direction. The temperature is maintained at $k_{\rm B}T=1$ using the Langevin thermostat for the relative velocity ${\bm v}-\dot{\gamma}y{\bm e}_{x}$.
The viscosity is calculated from the following equation:
\begin{eqnarray}
  \eta &=& \tau_{xy} / \dot{\gamma}, \label{eq:eta}
\end{eqnarray}
where $\tau_{xy}$ is the off-diagonal element of the virial stress. Here, $\dot{\gamma}$ ranges from $0.0005$ to $0.5$. The determined viscosities are shown in Fig.~\ref{fig:viscosity}. The reference liquid can be considered as a Newtonian fluid because the viscosity is almost constant as $\eta_{\rm ref}\simeq 3.5$. In contrast, the polymer solutions exhibit significant dependence, especially for solutions of longer polymers. Systems with both $N_{\rm s}=10$ and $100$ exhibit shear thinning, and the viscosity almost equals the viscosity of the reference liquid for $\dot{\gamma} \sim 0.1$. These viscosity dependences on the shear rate, concentration, and molecular weight are consistent with experiments, such as experiments on polystyrene solutions.~\cite{kk84}
\begin{figure}
\includegraphics{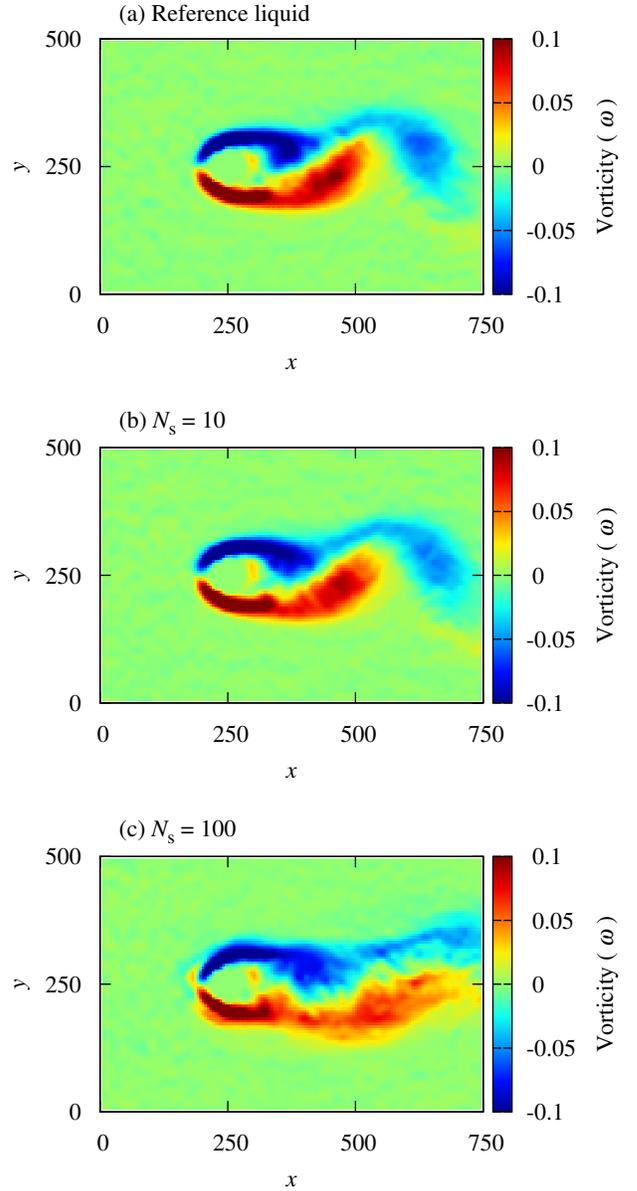}
\caption{\label{fig:vortex}Instantaneous vorticity plots at $Re=64$ for (a) reference liquid, (b) polymer solution with $N_{\rm s}=10$, and (c) polymer solution with $N_{\rm s}=100$. The mole fraction of the polymer is $\phi=0.107$.}
\end{figure}
\begin{figure}
\includegraphics{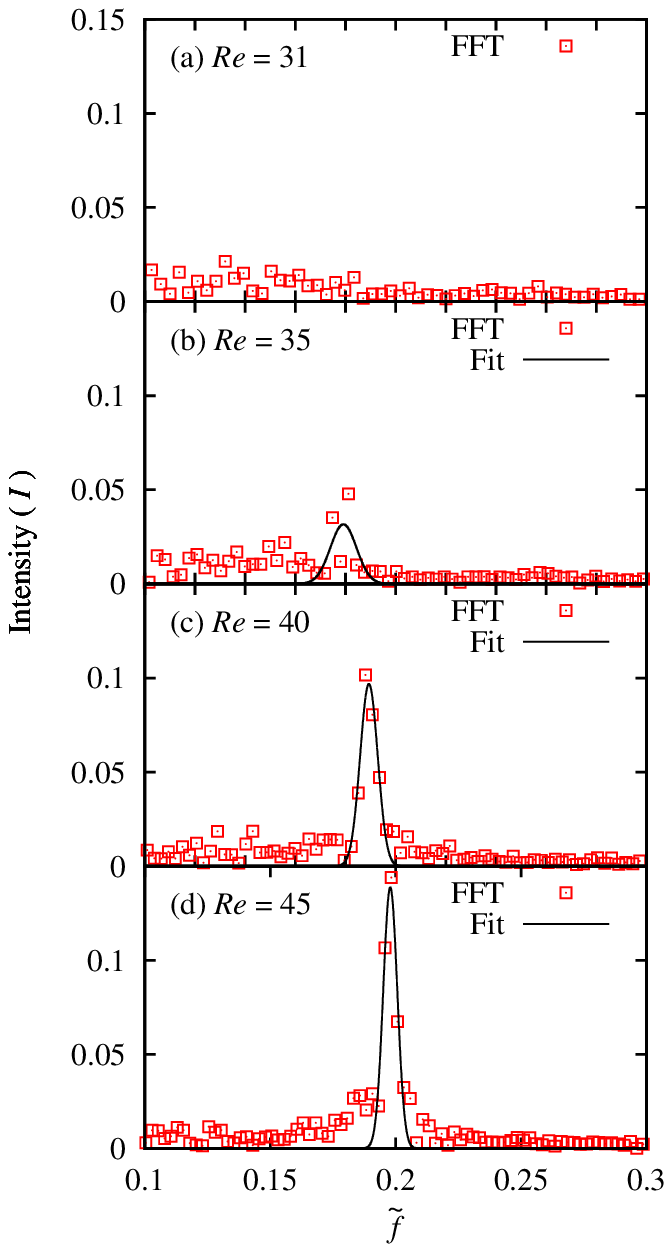}
\caption{\label{fig:fft_wca}Power spectra of the lift coefficient for the reference liquid. (a) Reynolds number $Re=31$, (b) $Re=35$, (c) $Re=40$, and (d) $Re=45$. The fitting results obtained by using Eq.~(\ref{eq:fit}) are also shown as solid lines.}
\end{figure}
\section{RESULTS}
\subsection{Typical vortex structures}
Typical vortex structures are shown in Fig.~\ref{fig:vortex}. To evaluate the vorticity fields, the computational domain is divided into square cells with side length $10$. The vorticity of each cell is then determined by the following equation:
\begin{eqnarray}
  \omega &=& \frac{ \partial {\bar{v}}_{y} }{ \partial x } - \frac{ \partial {\bar{v}}_{x} }{ \partial y }, \label{eq:vorticity}
\end{eqnarray}
where $\bar{v}_{\alpha}~(\alpha\in\{x,y\})$ represents the $\alpha$ component of the velocity of each cell averaged over $10~000$ steps. The central difference formula is adopted to compute the differentiation in Eq.~(\ref{eq:vorticity}).

As shown in Fig.~\ref{fig:vortex}, the vortex structure of the polymer solution for $N_{\rm s}=10$ is almost unchanged from that of the reference liquid, whereas for $N_{\rm s}=100$ the vortices are significantly blurred. In Secs. \ref{3b} and \ref{3c}, the effects of polymers on the vortex shedding are quantitatively described.
\begin{figure}
\includegraphics{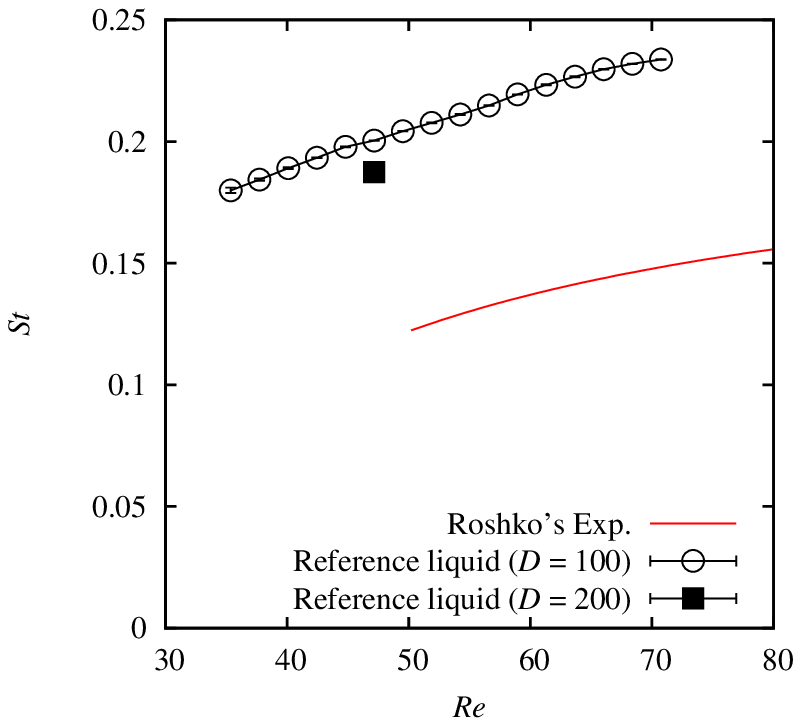}
\caption{\label{fig:st_wca}The Strouhal number $St$ as a function of the Reynolds number $Re$. The solid line denotes Roshko's experimental result~{\cite{roshko54}} given by Eq.~(\ref{eq:roshko}). The solid square shows the result of the large simulation with $D=200$.}
\end{figure}
\subsection{Reference liquid \label{3b}}
First, we study the properties of the liquid without polymers. The power spectra obtained from the Fourier transformation of the lift coefficient are shown in Fig.~\ref{fig:fft_wca}. The horizontal axis denotes a normalized frequency $\tilde{f}=fD/V$, where $f$ denotes the frequency of $C_{\rm L}$. The time series of the lift coefficient is recorded every $10$ time steps, and the total number of the time series data is $2^{19}$. Although no peak can be found in the spectrum of $Re=31$, characteristic peaks appear for $Re\ge 35$. Figure~\ref{fig:fft_wca} also shows that the peak position depends on $Re$. To evaluate the Strouhal number $St$ given by Eq.~(\ref{eq:st}), the intensity $I(\tilde{f})$ of the power spectrum is fitted by the Gaussian function:
\begin{eqnarray}
  I\left(\tilde{f}\right) &=& A{\rm exp}\left(-\frac{\left(\tilde{f}-St\right)^2}{2\sigma_{St}^2}\right), \label{eq:fit}
\end{eqnarray}
where $A$ and $\sigma_{\rm St}$ are the fitting parameters. The solid lines in Fig.~\ref{fig:fft_wca} denote the fitting results. The averages and errors of $St$ and the fitting parameters are calculated from eight independent runs from different initial conformations.

Figure~\ref{fig:st_wca} shows the $Re$ dependence of $St$. In the case of Newtonian fluid flow past a circular cylinder, the relation between $St$ and $Re$ is established by the following empirical formula~\cite{roshko54}:
\begin{eqnarray}
  St&=&0.212-\frac{4.5}{Re}.\label{eq:roshko}
\end{eqnarray}
Our results for the reference liquid are qualitatively similar but approximately twice as high as those of the experiments. This discrepancy between our simulation and the experiments is likely caused by the finite-size effect of the simulation, because the system is very small compared to that of the experiments. We discuss details in the Appendix~\ref{app}. In the present study, we discuss the properties of the polymer solutions on the basis of the reference liquid.
\begin{figure}
\includegraphics{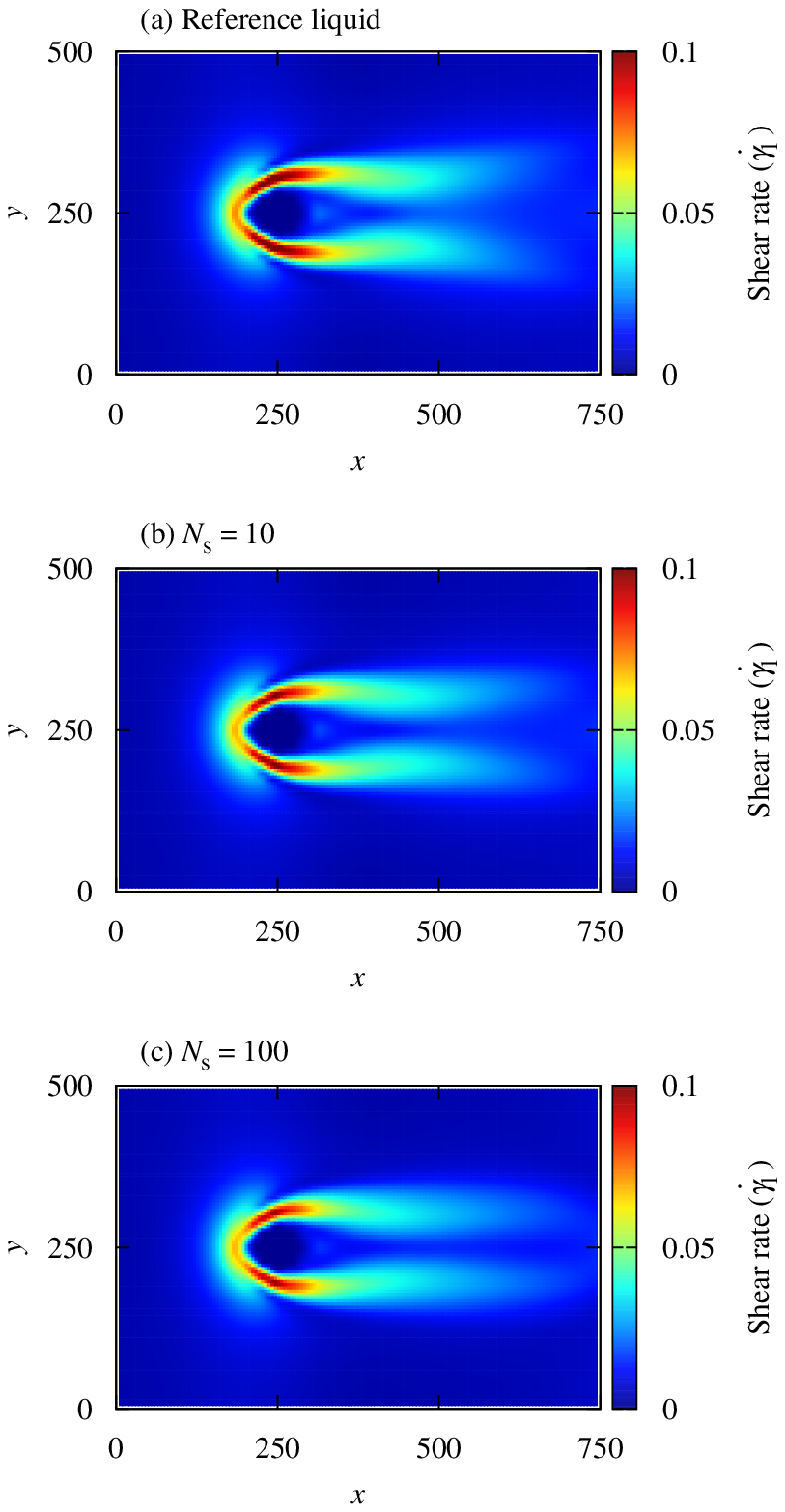}
\caption{\label{fig:shear}Shear rate fields for the fluid at $V=1.9$: (a) reference liquid, (b) polymer solution with $N_{\rm s}=10$,  and (c) polymer solution with $N_{\rm s}=100$. The mole fraction of the polymer solutions is $\phi=0.107$.}
\end{figure}
\begin{figure}
\includegraphics{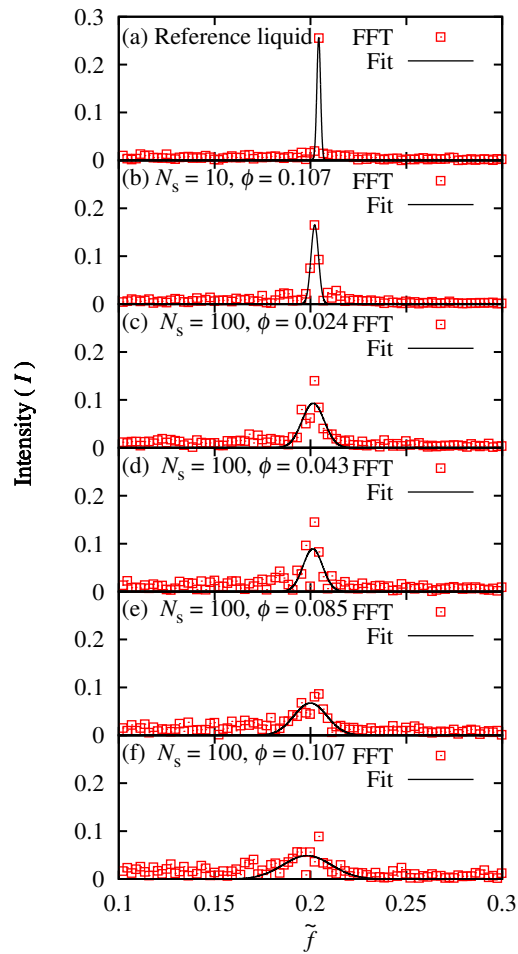}
\caption{\label{fig:fft_polymer}Power spectrum of the lift coefficient of the obstacle at $Re=50$: (a) reference liquid, (b) polymer solution ($N_{\rm s}=10$, $\phi=0.107$), (c) polymer solution ($N_{\rm s}=100$, $\phi=0.024$), (d) polymer solution ($N_{\rm s}=100$, $\phi=0.043$), (e) polymer solution ($N_{\rm s}=100$, $\phi=0.085$), and (f) polymer solution ($N_{\rm s}=100$, $\phi=0.107$). The fitting results by using Eq.~(\ref{eq:fit}) are shown as solid lines.}
\end{figure}
\subsection{Polymer solution \label{3c}}
To determine the Reynolds number, the value of the viscosity is required.
However, it is difficult to determine this, because the viscosity of the polymer solution strongly depends on the shear rate as shown in Fig.~\ref{fig:viscosity}. 
In order to consider the characteristic value of the viscosity, we first calculate the fields of the shear rate of each fluid.
The shear rates are estimated by the eigenvalue of the strain tensor:
\begin{eqnarray}
  S_{\alpha \beta}=\frac{1}{2}\left(\frac{\partial {\bar{v}}_{\beta}}{\partial \alpha} + \frac{\partial {\bar{v}}_{\alpha}}{\partial \beta} \right), \label{eq:shear}
\end{eqnarray}
where $\alpha,~\beta\in\{x,~y\}$.
The local shear rate $\dot{\gamma}_{\rm l}$ is estimated as follows:
\begin{eqnarray}
  \dot{{\gamma}}_{\rm l} &=& |\lambda_{1} - \lambda_{2}|,\label{eq:shear_rate} 
\end{eqnarray}
where $\lambda_{1}$ and $\lambda_{2}$ are the eigenvalues of the strain tensor.
To evaluate the space derivatives of the velocity field, we adopt the same method used in Eq.~(\ref{eq:vorticity}). The distribution of the time-averaged shear rates of each fluid for $V=1.9$ is shown in Fig.~\ref{fig:shear}. 
Since the shear rates are inhomogeneous, the value of the viscosity varies from place to place.
However, the physics of vortex shedding is likely governed by the flow near the cylinder.
When the inlet velocity $V$ is strong ($V \ge 1.9$), then the shear rate near the cylinder is higher than $0.1$.
In this region, the value of the viscosity of the polymer solution is very close to that of the reference liquid, as shown in Fig.~\ref{fig:viscosity}. Therefore, we employ the reference liquid viscosity $\eta_{\rm ref}$ to determine the value of the Reynolds number of polymer solutions.

\begin{figure}
\includegraphics{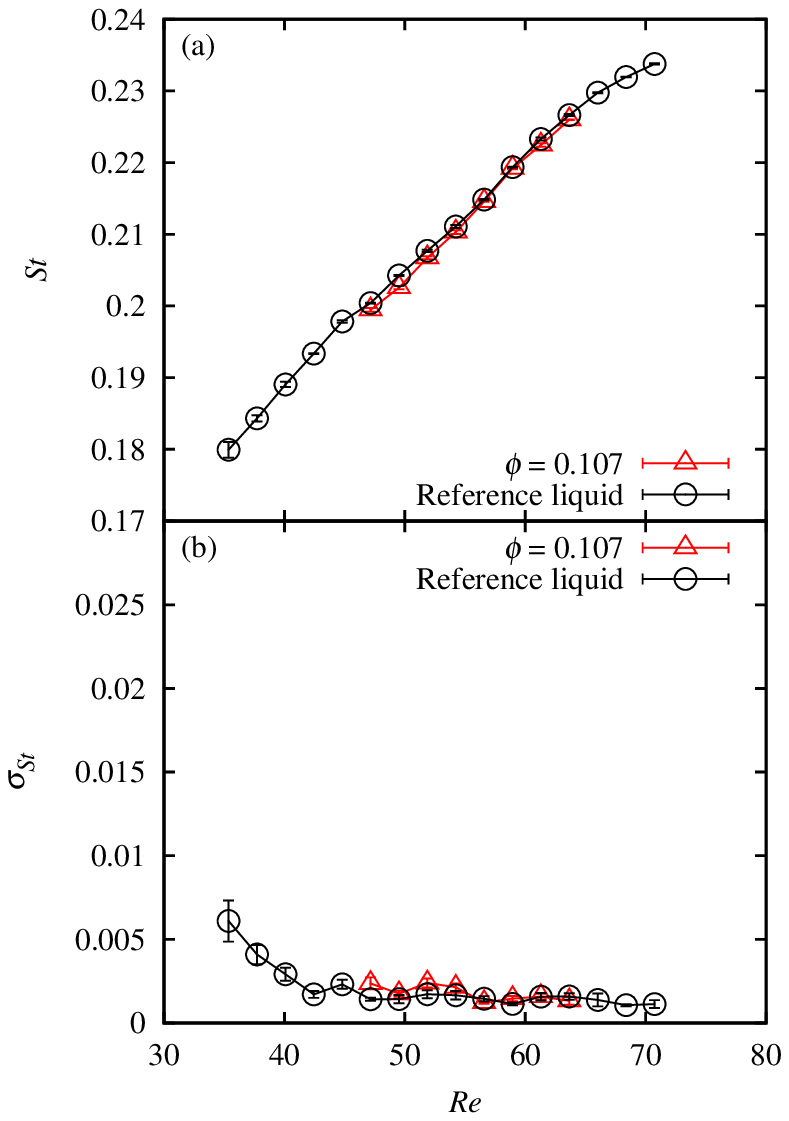}
\caption{\label{fig:st_ns10}(a) The Strouhal number $St$ and (b) the spectral width $\sigma_{St}$ as a function of the Reynolds number $Re$. The circles denote the results of the reference liquid. The results of polymer solution with $N_{\rm s}=10$ are shown for mole fraction $\phi=0.107$.}
\end{figure}

\begin{figure}
\includegraphics{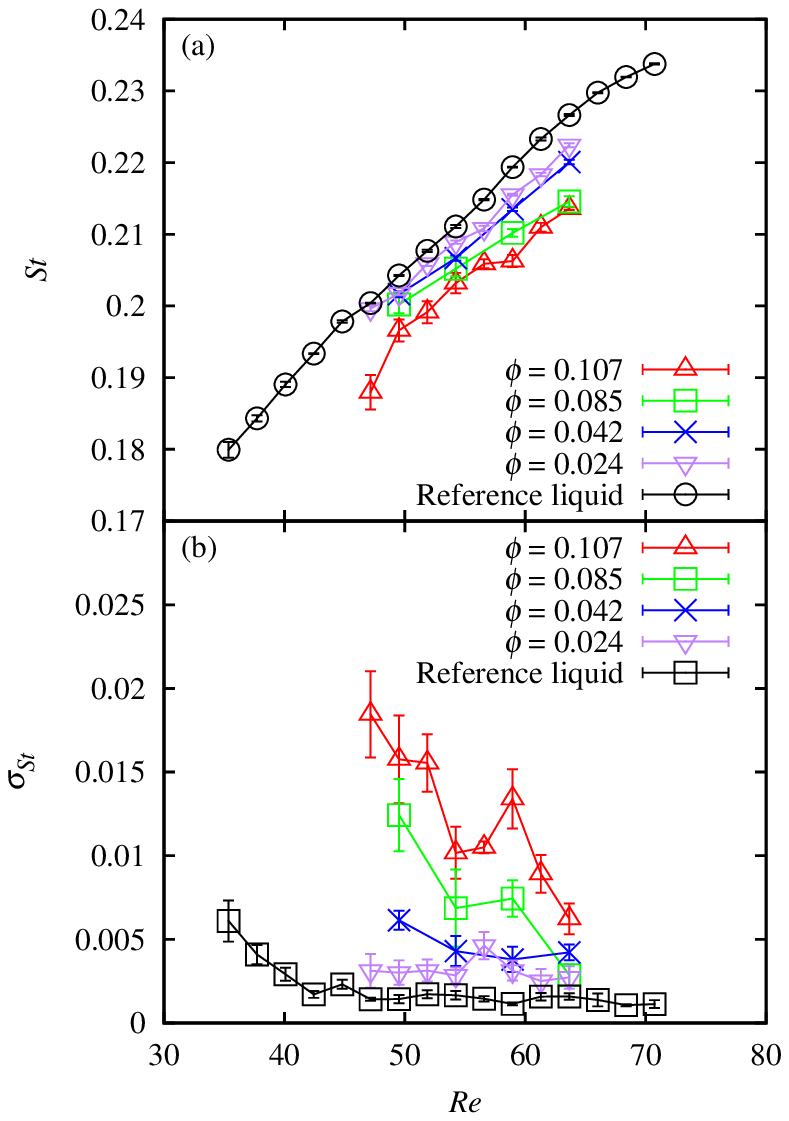}
\caption{\label{fig:st_ns100}(a) The Strouhal number $St$ and (b) the spectral width $\sigma_{St}$ as a function of the Reynolds number $Re$.  The circles denote the results for the reference liquid. The results of polymer solutions with $N_{\rm s}=100$ are shown for mole fraction $\phi=0.024$, $0.043$, $0.085$, and $0.107$.}
\end{figure}

The power spectra of the lift coefficient at $Re=50$ are shown in Fig.~\ref{fig:fft_polymer}. 
While there is no significant difference between the spectra of the polymer solution with $N_{\rm s}=10$ and the reference liquid, the spectra with $N_{\rm s}=100$ exhibit an apparent dependence on the mole fraction $\phi$. As $\phi$ increases, the peak position shifts to the lower frequency, and the peak is broadened. In order to confirm these results quantitatively, we determine the peak position and the width of the spectrum by fitting Eq.~(\ref{eq:fit}) and estimate the value of $St$ from the determined peak position as shown in Figs.~{\ref{fig:st_ns10}} and \ref{fig:st_ns100}. 
The Reynolds number dependences of the Strouhal number $St$ and the width of the spectrum of the short-polymer solution ($N_{\rm s}=10$)
exhibit no significant difference compared to those of the reference liquid (see Fig.~{\ref{fig:st_ns10}}).
However, the behavior of the long-polymer solution ($N_{\rm s}=100$) significantly deviates from that of the reference liquid (see Fig.~{\ref{fig:st_ns100}}). As $\phi$ increases, the shedding frequency decreases, and the width of the peak increases. These behaviors are consistent with the experimental results.
\begin{figure}
\includegraphics{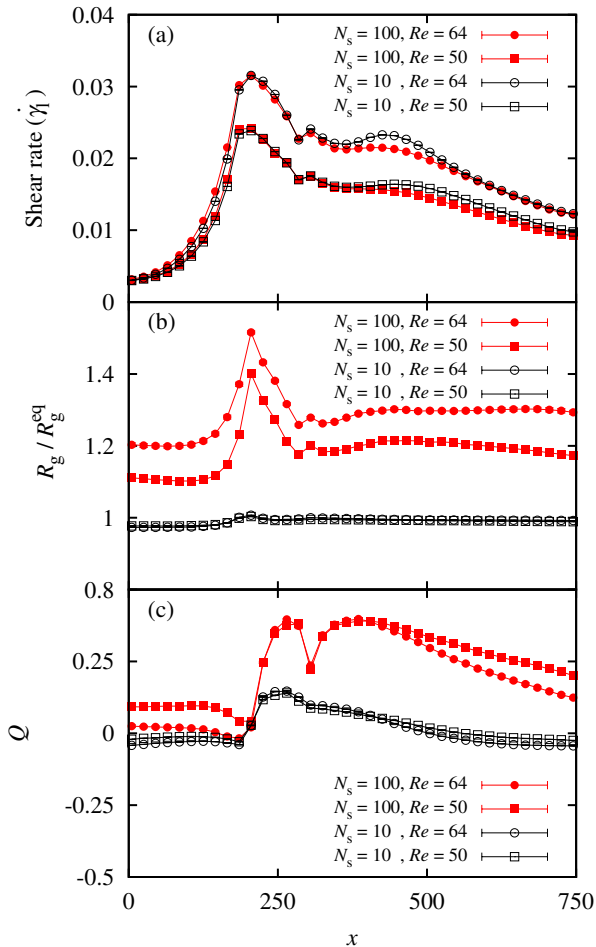}
\caption{\label{fig:y_average}(a) The local shear rate $\dot{\gamma}_{\rm l}$, (b) the gyration radius $R_{\rm g}$, and (c) the orientational order parameter $Q$ for the short-polymer solution ($N_{\rm s}=10$) and the long-polymer solution ($N_{\rm s}=100$) as a function of $x$. All quantities are averaged along the $y$-axis. The mole fraction of the polymer solution is $\phi=0.107$.}
\end{figure}
\begin{figure}
\includegraphics{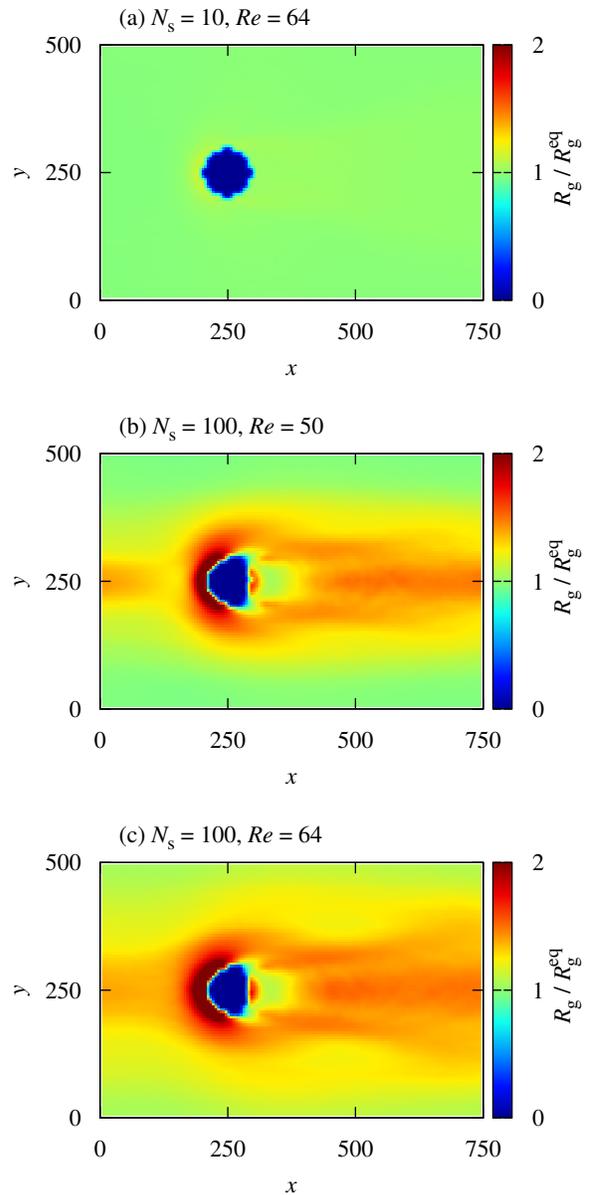}
\caption{\label{fig:rg}Distributions of gyration radius $R_{\rm g}$ at $\phi=0.107$ for (a) the short-polymer solution ($N_{\rm s}=10$) with Reynolds number $Re=64$ and the long-polymer solution ($N_{\rm s}=100$) with (b) $Re=50$ and (c) $Re=64$.
}
\end{figure}
\begin{figure}
\includegraphics{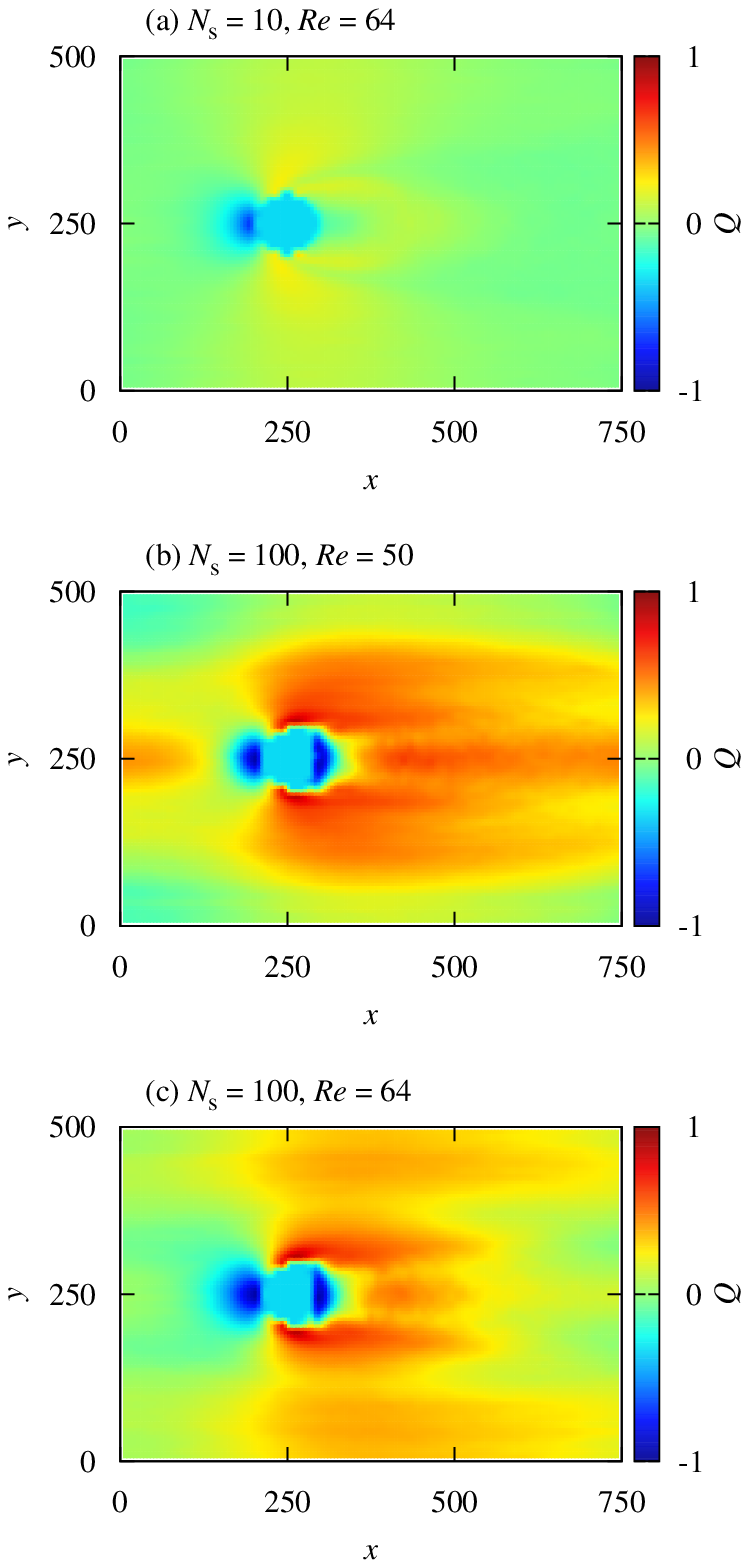}
\caption{\label{fig:dop}Distributions of the orientational order parameter $Q$ at $\phi=0.107$
for (a) the short-polymer solution ($N_{\rm s}=10$) with Reynolds number $Re=64$ and the long-polymer solution ($N_{\rm s}=100$) with (b) $Re=50$ and (c) $Re=64$.}
\end{figure}
\section{DISCUSSION}
Figures~{\ref{fig:st_ns10}} and {\ref{fig:st_ns100}} show that the long polymers suppress the vortex shedding, whereas the short polymers do not. Therefore, the extensibility of the polymer plays an important role in the flow.
First, we consider the polymer effect on the shear rate, because the vortex shedding cycle is related to the shear stress over a wide area of the near wake.~\cite{gg93}
The distribution of the time- and space-averaged shear rates of each fluid is shown in Fig.~{\ref{fig:y_average}}(a).
The space average is taken over the $y$ direction.
Despite there being no significant difference between the shear rates of the long-polymer solution and short-polymer solution in the upstream region, the shear rates of the long-polymer solution are lower than those of the short-polymer solution in the wake region. This decrease in the shear rate is due to the stretching of the polymer, because the strain energy is consumed by the extension of the polymer.

In order to evaluate the stretching of the polymer, we consider the gyration radius.
The gyration radius is given by the following equation:
\begin{eqnarray}
  R_{\rm g}^2 &=& \frac{1}{N_{\rm s}}\sum_{i=1}^{N_{\rm s}}\left\langle \left({\bm{R}}_i - {\bm{R}}_{\rm G} \right)^2\right\rangle,
\end{eqnarray}
where ${\bm{R}}_i$ denotes the position vector of $n$th segment of a polymer, and ${\bm R}_{\rm G}$ is the center of mass of the polymer, respectively. The angular brackets denote the time average at each cell.
In thermal equilibrium, the gyration radius of the short and long polymers are $R_{\rm g}^{\rm eq}=1.8$ and 10.2, respectively.
As shown in Figs.~{\ref{fig:y_average}}(b) and {\ref{fig:rg}}(a), the gyration radii of the short polymers are almost uniform and are almost identical to the equilibrium value. Therefore, the short polymers behave like particles in the fluid. On the other hand, the long polymers are strongly stretched in flows as shown in Figs.~{\ref{fig:y_average}}(b), {\ref{fig:rg}}(b), and {\ref{fig:rg}}(c). Therefore, the long polymers change their shapes in the flow, and this affects the behavior of the vortex shedding.

When the flow is slow ($Re=50$), the gyration radius of the long polymer gradually decreases as $x$ increases, because the shear rate decreases. However, the gyration radius is almost constant for the fast flow ($Re=64$). This behavior is likely to be due to the difference in strength of the vortices. In the latter case, some polymers in the near wake region are caught up in vortices.
In order to evaluate the relation between the polymer motion and the strength of the flow field, we also investigate the orientation of the polymers.
The orientational order parameter is given by the following equation:
\begin{eqnarray}
  Q&=& 2\left\langle \cos^2 \theta \right\rangle - 1,
\end{eqnarray}
where $\theta$ denotes the angle between the direction of the end-to-end vector of each polymer and the $x$-axis. The order parameter becomes unity when all the polymers align in the flow direction, and it becomes $-1$ when all the polymers align perpendicular to the flow.
When the polymers are randomly directed, the order parameter becomes zero.
The order parameter of the short-polymer solution is almost homogeneous and is close to zero (see Figs.~{\ref{fig:y_average}}(c) and {\ref{fig:dop}}(a)). This result is reasonable because each short polymer forms a ball and has a random orientation in the flow. The order parameters of the long-polymer solutions are highly inhomogeneous as shown in Figs.~{\ref{fig:y_average}}(c), {\ref{fig:dop}}(b), and {\ref{fig:dop}}(c). In the slow flow case ($Re=50$), the polymers behind the cylinder are stretched and orient to the flow direction. However, the polymers behind the cylinder orient homogeneously in the fast flow case ($Re = 64$). This suggests that the polymers in the fast flow case are caught up in vortices. Therefore, the suppression of the vortex shedding is related to not only the stretching of the polymer by the shear but also to the entrainment of the polymer by the vortices.  As a consequence, the shapes of the vortices become blurred.

\section{SUMMARY}
We have studied the effects of polymers on the flow around a cylinder at the molecular scale by using MD simulations. 
When the polymers are short, the behavior of the polymer solution is almost identical to that of the liquid without polymers.
However, the flow behavior significantly changes when long polymers are added.
This means that the extensibility of polymers strongly affects the flow patterns, as suggested by Cressman~{\it{et~al.}}~\cite{cbg01}
By observing the power spectra of the force acting on the cylinder, we found that a peak in the spectrum is broadened as the mole fraction of the long polymers increases. This quantifies the blurring of the shape of the vortices for polymer solutions. The shift and broadening of the peaks in the spectra are consistent with the results of the experiments. We conclude that the suppression of the vortex shedding is related to not only the shear stress in the wake region but also to the strength of the vortices.

Unlike the simulation on the basis of the Navier--Stokes equation, 
MD simulation allows us to obtain information at the molecular level.
We show that the shape deformation of polymer chains plays a crucial role in the blurring of the K\'arm\'an vortices.
Our results show that MD simulations are a useful tool to explore other complex flows involving molecular aggregates and nanometer-scale objects such as surfactants and nanobubbles.

\begin{acknowledgments}
We would like to thank Toshihiro Kawakatsu, Youhei Morii, and Yuji Higuchi  for helpful discussions. This research was supported by MEXT as ``Exploratory Challenge on Post-K computer'' (Challenge of Basic Science -- Exploring Extremes through Multi--Physics and Multi--Scale Simulations) and JSPS KAKENHI Grant Numbers JP15K05201. Computation was partially carried out by using the facilities of the Supercomputer Center, Institute for Solid State Physics, University of Tokyo. 
\end{acknowledgments}

\begin{figure}
\includegraphics{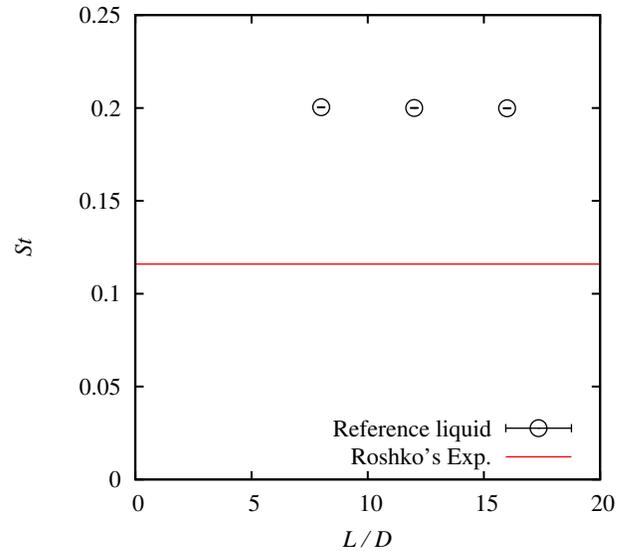}
\caption{\label{fig:change_Lx}The Strouhal number $St$ as a function of the length of the channel $L=L_x - L_{\rm t}$ at the Reynolds number $Re=47$ for the reference liquid ($\phi=0$), where $L_{\rm t}$ is the width of the thermostat region.}
\end{figure}

\begin{figure}
\includegraphics{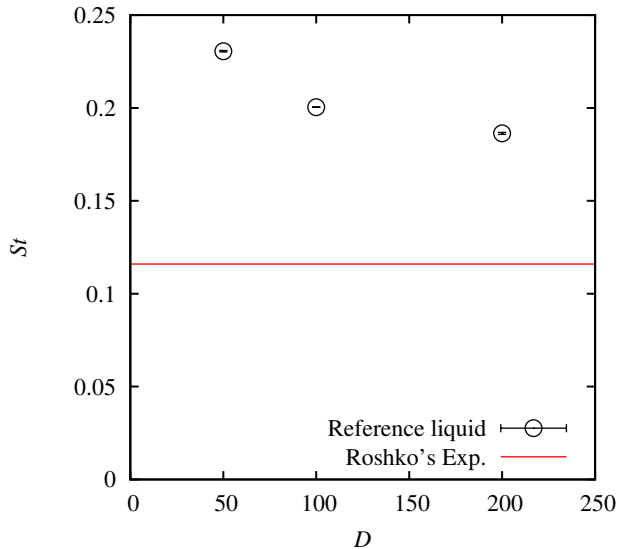}
\caption{\label{fig:change_Dsize}The Strouhal number $St$ as a function of the cylinder diameter $D$ at the Reynolds number $Re=47$ for the reference liquid ($\phi=0$).}
\end{figure}

\appendix
\section{SYSTEM SIZE EFFECTS\label{app}}
As shown in Fig.~\ref{fig:st_wca}, the values of $St$ differ from the experimental data by a factor $2$ in Newtonian fluids. We consider that this discrepancy is due to the finite-size effect. When a system is small, the compressibility of liquid is not negligible, because a high velocity is necessary to achieve vortex shedding, whereas the liquid used in the experiments can be considered incompressible. It is known that the Mach number and temperature gradient affect $St$.~\cite{twt14, ouv06, sd03} In addition, the vortices interfere with each other in a small system due to the periodic boundary condition. The interference also affects $St$.~\cite{swp99} In order to investigate the finite-size effect on $St$, we investigate the channel length and the compressibility dependence of $St$. The channel length $L$ is defined as $L=L_x-L_{\rm t}$, where $L_{\rm t}$ is the width in the $x$ direction of the thermostat region. 

In order to investigate the $L$ dependence of $St$, we estimate the $St$ with varying $L$ keeping the other parameters constant. Figure~\ref{fig:change_Lx} shows the $L/D$ dependence of $St$ at $Re=47$. The channel length has no effect on $St$. Therefore, we conclude that the channel length of $L/D=8$ which adopted in the present study is long enough for eliminating the finite-size effect in the flow direction.
As for the effect of the system size compared to the particle diameter (flow resolution), we estimate the $St$ with varying the cylinder diameter $D$ keeping $L/D=8$. Figure~\ref{fig:change_Dsize} shows the $D$ dependence of $St$ at $Re=47$. Since the inlet velocity $V$ decreases as $V \sim 1/D$ at a constant $Re$, the temperature gradient caused by the flow and the Mach number $Ma$ decrease: $Ma=0.67, 0.33$ and $0.17$ at $D=50, 100$ and $200$, respectively.
The solid square in Fig.~\ref{fig:st_wca} depicts the results for the largest system size in this investigation. The system size is $(L_x, L_y) = (2000, 1000)$ and cylinder diameter $D=200$ whose center is located at $(x,y)=(500, 500)$.
Although the system size has slight effects on $St$ as shown in Fig.~\ref{fig:change_Dsize}, the change is small compared to the increase of the computational cost.

We also investigate the effects of the periodic boundary condition. We performed simulations with increasing $L_y$ while keeping the length of $L_x$ constant. Although the value of $St$ tends to decrease as $L_y$ increases, the change is small. The amount of decrease for the largest system size in this investigation is about $6\%$ with the size of $(L_x, L_y) = (1000, 2000)$, which is almost same as the solid square in Fig.~\ref{fig:st_wca}.
Therefore, we conclude the main reason is the finite size effects of the cylinder and the channel width. Further investigations are necessary to quantitatively address this problem. Since the main objective is the investigation of the polymer effects on the K\'arm\'an vortex in a wide range of parameter space, we adopt the simulation box shown in Fig.~\ref{fig:simulation_box}.

\providecommand{\noopsort}[1]{}\providecommand{\singleletter}[1]{#1}%

\end{document}